**Comments on "Assessing future risk: quantifying the effects of sea level rise on storm surge risk for the southern shores of Long Island, New York," by Christine C. Shepard, Vera N. Agostini, Ben Gilmer, Tashya Allen, Jeff Stone, William Brooks and Michael W. Beck** (*Natural Hazards,* Vol. 60, No. 2, 727-745, DOI: 10.1007/s11069-011-0046-8)

by David A. Burton

**Abstract:** Tide gauge and satellite data indicate that the rate of sea level rise has not increased significantly in response to the last 3/4 century of $CO_2$ emissions, so there is no reason to expect that it will do so in response to the next 3/4 century of $CO_2$ emissions. The best prediction for sea level in the future is simply a linear projection of the history of sea level at the same location in the past. For Long Island, that is about 7-8 inches by 2080.

At its western tip, Long Island meets Manhattan Island, at a place called The Battery. There is an excellent GLOSS-LTT tide gauge there which has been measuring sea levels since 1856.

Due to local land subsidence, sea level is rising faster at The Battery than at 85% of the other GLOSS-LTT tide gauges in the world, but the rate of rise has been nearly constant for over a century, at 2.77 +/- 0.09 mm/year (95% confidence interval).

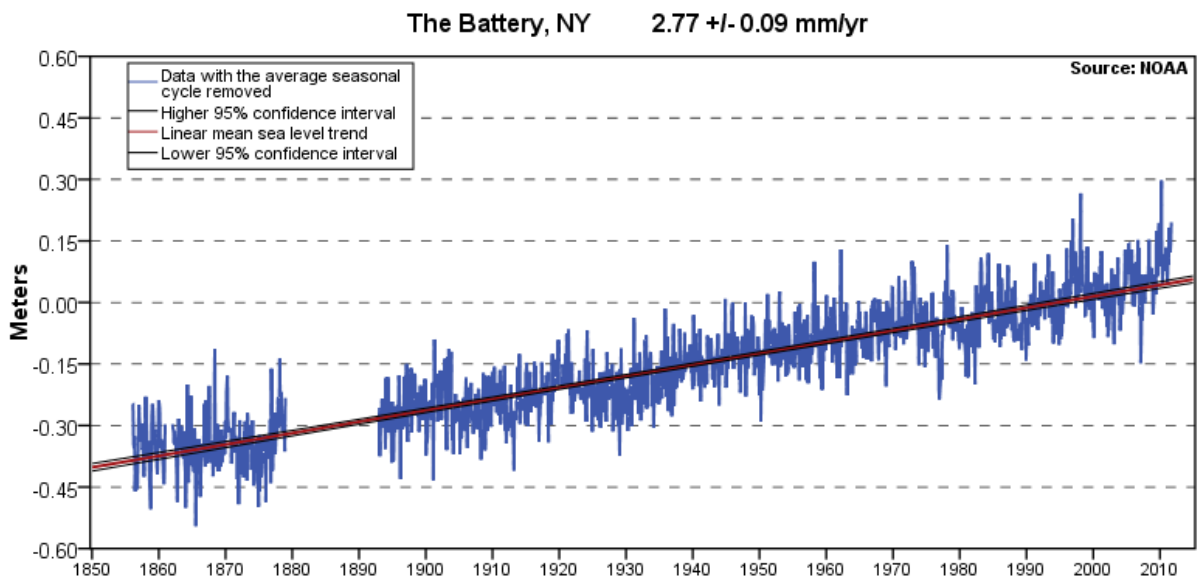

About 32 Km away, at Kings Point, on the north shore of Long Island, there's another good GLOSS-LTT tide gauge, which has been measuring sea level since 1931. Sea level there has been rising at only 2.35 +/- 0.24 mm/year.

Despite over 2/3 century of major anthropogenic $CO_2$ emissions, these tide gauges have measured no statistically significant increase in the rate of sea level rise.

The same thing is true at most other tide gauges around the world. In fact, the best and most comprehensive analyses of sea level measured by tide gauges around the world show slight decelerations in the rate of sea level rise over the last 80 years. [*Houston, J.R. and Dean, R.G., 2011*. *"Sea-Level Acceleration Based on U.S. Tide Gauges and Extensions of Previous Global-Gauge Analyses."* Journal of Coastal Research, 27(3), 409–417, DOI: 10.2112/JCOASTRES-D-10-00157.1] (Note: sea level isn't rising everywhere; at some locations sea level is falling.)

Another way of measuring sea level is via satellite altimeters, which measure the height of the sea surface over the open ocean. We only have about 19 years of data from the satellites, so far, which is too short to draw definite conclusions, but they, too, are showing deceleration in the rate of sea level rise.

If we project the high end of the range of sea level rise for The Battery to 2080 (2.77+0.09 = 2.86 mm/year x 69 years), it adds up to just 20 cm (7.8 inches). There is no reason to expect that sea level will rise more than that by 2080.

Even the IPCC Fourth Assessment Report only projected 7 - 23 inches (15" mid-range) average sea level rise for 100 years. Yet Shepard, *et al,* postulate 20 inches of sea level rise for a shorter 69 year period (and call it "modest and probable"), based on models which indicate that sea level rise *should be* accelerating in response to higher atmospheric $CO_2$ levels. But the data indicate that those models are wrong, sea level rise is not accelerating.

A "modest and probable" amount of sea level rise by 2080 for New York and Long Island is about 7 inches, not 20 inches.

The IPCC's Third Assessment Report noted the *"observational finding of no acceleration in sea level rise during the 20th century."* But there is still a lot of confusion about sea level rise. Much of it results from misunderstanding the findings of a key paper, Church, J. A., and White, N. J., 2006, *"A 20th Century Acceleration in Global Sea-Level Rise."* Geophysical Research Letters, Vol. 33, L01602, 4 PP. DOI: 10.1029/2005GL024826

Church & White fit a quadratic to averaged and adjusted tide gauge data, and detected a small acceleration in rate of sea level rise for the 20th century as a whole. But it turns out that all of that acceleration occurred in the first quarter of the 20th century (and the late 19th century). After 1925, their data showed a small deceleration in rate of sea level rise, rather than acceleration.

Since nearly all of the anthropogenic contribution to $CO_2$ levels occurred after 1925, that means Church & White detected no acceleration in rate of sea level rise in response to anthropogenic $CO_2$.

In 2009, Church and White released a new data set, based on a different set of tide gauges. I applied their 2006 analysis method to the new data. I found that it not only showed deceleration in sea level rise after 1925, all of the acceleration in sea level rise for the full 20th century was also gone. I shared my results with Drs. Church & White, and on June 18, 2010, Dr. Church replied, confirming my analysis: *"For the 1901 to 2007 period, again we agree with your result and get a non-significant and small deceleration."*

In 2011, Church and White released a third data set. This one shows a very slight acceleration in sea level rise after 1925, though much smaller in magnitude than the deceleration seen in their other data sets. The post-1925 acceleration in this data set, if it continued to 2080, would add just 0.8 inches of sea level rise, compared to a linear projection.

Since the rate of sea level rise has not increased significantly in response to the last 3/4 century of $CO_2$ emissions, there is no reason to expect that it will do so in response to the next 3/4 century of $CO_2$ emissions. The best prediction for sea level in the future is simply a linear projection of the history of sea level at the same location in the past, or about 7-8 inches by 2080, for Long Island.

However, all is not lost for this report. Long Island will probably experience 20 inches of sea level rise by about 2200. So if they just change "2080" to "2200," the report will be useful for projecting the impact of sea level rise on Long Island residents 190 years from now.

Note: The graphs, code and data alluded to in this comment can be viewed here:
**http://tinyurl.com/nhazburt1**

David A. Burton
Cary, NC  USA
Member, North Carolina Sea Level Rise Impact Study Advisory Committee
Expert Reviewer, IPCC AR5 WG1 FOD
http://www.burtonsys.com/email/

**References:**

1. Christine C. Shepard, Vera N. Agostini, Ben Gilmer, Tashya Allen, Jeff Stone, William Brooks and Michael W. Beck. Assessing future risk: quantifying the effects of sea level rise on storm surge risk for the southern shores of Long Island, New York. *Natural Hazards,*, Volume 60, Number 2, 727-745. DOI: 10.1007/s11069-011-0046-8

2. James R. Houston and Robert G. Dean. Sea-Level Acceleration Based on U.S. Tide Gauges and Extensions of Previous Global-Gauge Analyses. *Journal of Coastal Research*, 27(3), 409–417. DOI: 10.2112/JCOASTRES-D-10-00157.1

3. John A. Church and Neil J. White. A 20[th] Century Acceleration in Global Sea-Level Rise. *Geophysical Research Letters*, Vol. 33, L01602, 4 pp. DOI: 10.1029/2005GL024826